\documentclass[lettersize,journal]{IEEEtran}
\usepackage{subfigure}
\usepackage{enumitem}

\usepackage[utf8]{inputenc} 
\usepackage[T1]{fontenc}    
\usepackage{url}            
\usepackage{booktabs}       
\usepackage{amsfonts}       
\usepackage{nicefrac}       
\usepackage{microtype}      
\usepackage{xcolor}         
\usepackage[utf8]{inputenc} 
\usepackage[T1]{fontenc}    
\usepackage{url}            
\usepackage{booktabs}       
\usepackage{amsfonts}       
\usepackage{nicefrac}       
\usepackage{microtype}      
\usepackage{xcolor}     
\usepackage{algorithmic}
\usepackage{amsmath}
\usepackage{graphicx}
\usepackage{epstopdf}
\usepackage[ruled,linesnumbered]{algorithm2e}
\usepackage{multirow} 
\usepackage{tabularx}
\usepackage{makecell}
\newcolumntype{P}[1]{>{\centering\arraybackslash}p{#1}}
\newcolumntype{M}[1]{>{\centering\arraybackslash}m{#1}}

\usepackage{orcidlink}

\AtBeginDocument{%
  \providecommand\BibTeX{{%
    \normalfont B\kern-0.5em{\scshape i\kern-0.25em b}\kern-0.8em\TeX}}}

\begin{document}

\def\oursys{DeepCore\xspace}
\title{\oursys: Simple Fingerprint Construction for Differentiating Homologous and Piracy Models}

\author{Haifeng Sun, Lan Zhang and Xiang-Yang Li
\thanks{University of Science and Technology of China; Hefei, China; Email: sun1998@mail.ustc.edu.cn, zhanglan@ustc.edu.cn, xiangyangli@ustc.edu.cn}}

\markboth{Journal of \LaTeX\ Class Files,~Vol.~14, No.~8, August~2021}%
{Shell \MakeLowercase{\textit{et al.}}: A Sample Article Using IEEEtran.cls for IEEE Journals}


\maketitle

\begin{abstract}

As intellectual property rights, the copyright protection of deep models is becoming increasingly important. Existing work has made many attempts at model watermarking and fingerprinting, but they have ignored homologous models trained with similar structures or training datasets. We highlight challenges in efficiently querying black-box piracy models to protect model copyrights without misidentifying homologous models. To address these challenges, we propose a novel method called \oursys, which discovers that the classification confidence of the model is positively correlated with the distance of the predicted sample from the model decision boundary and piracy models behave more similarly at high-confidence classified sample points. Then \oursys constructs core points far away from the decision boundary via optimizing the predicted confidence of a few sample points and leverages behavioral discrepancies between piracy and homologous models to identify piracy models.
Finally, we design different model identification methods, including two similarity-based methods and a clustering-based method to identify piracy models using models' predictions of core points. Extensive experiments show the effectiveness of \oursys in identifying various piracy models, achieving lower missed and false identification rates, and outperforming state-of-the-art methods. 

\end{abstract}

\begin{IEEEkeywords}
Model Copyright Protection, Piracy Model, Homologous Model
\end{IEEEkeywords}

\section{Introduction}

In recent years, deep learning has witnessed rapid development and found extensive applications in various fields, such as computer vision \cite{taigman2014deepface}, speech recognition \cite{hoy2018alexa}, and natural language processing \cite{vaswani2017attention}. 
Many companies choose not to open-source deep learning models to protect their commercial interests. This is due to the significant resources required for training advanced neural network models, including massive datasets, substantial computing power, and the expertise of the designers. For example, training models like GPT-3 demand 45TB of data and incur training costs exceeding 12 million US dollars. 
However, the issue of model copyright faces numerous security threats. Adversaries could obtain white-box models through unconventional means and make modifications \cite{liu2018fine,molchanov2019importance} like fine-tuning, pruning, and adversarial training. Additionally, adversaries can steal models through model extraction attacks \cite{jagielski2020high,orekondy2019knockoff,zanella2021grey,chandrasekaran2020exploring}. Consequently, there has been an upsurge in research to address these threats~\cite{10143370}. Existing studies can be broadly categorized into two types. The first type employs model watermarking methods \cite{jia2021entangled,uchida2017embedding,fan2019rethinking,zhang2020passport,adi2018turning,ge2021anti}, which require modifying the original model and often lead to model performance degradation. Additionally, most watermarking methods struggle to withstand model extraction attacks. The second type adopts model fingerprinting methods \cite{lukas2019deep,cao2021ipguard,li2021modeldiff,guan2022you,peng2022fingerprinting}, with current research primarily focusing on decision boundaries. These methods characterize the similarity of decision boundaries using adversarial examples. However, adversaries can manipulate decision boundaries through adversarial training, rendering the fingerprint ineffective. 
\begin{figure}
  \centering
  \includegraphics[width=0.48\textwidth]{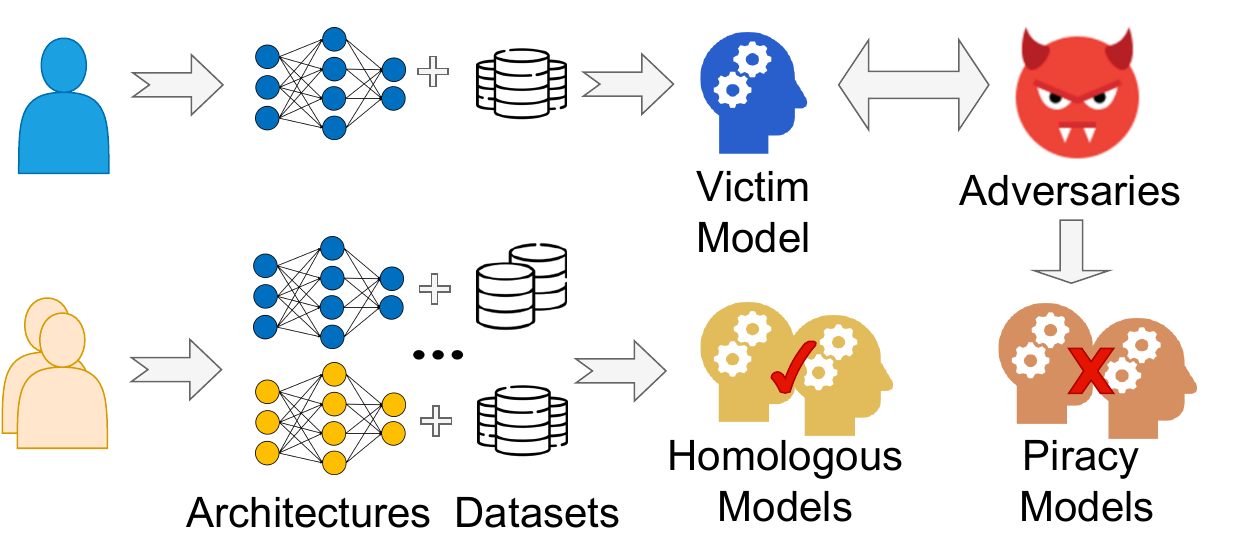}
  \caption{ Homologous models have similar model architectures or training data of the victim model and train independently. Piracy models depend on the victim model through illegal theft of the white-box model or model extraction attacks.}
  \label{scenario}
\end{figure}
Besides, as shown in figure~\ref{scenario}, there are homologous models trained by other legitimate users using similar model architectures or datasets. The existence of homologous models could increase the difficulty of identifying piracy models. However, existing work ignored the study of distinguishing homologous models.
Consequently, there is an urgent need to safeguard deep neural network models against illegal copying and protect homologous models.

Our research confronts three main challenges. Firstly, we strive to ensure no piracy models are overlooked and avoid misidentifying homologous models. This cannot be easy because similar model architectures and datasets could train similar models. 
Secondly, the model owner only has black-box access to piracy models and gets the models' predictions of query samples. 
Thirdly, the model fingerprint not only needs to be effective but also efficient. Using as few query samples as possible can reduce costs and, at the same time, avoid being detected by adversaries. 
To address these challenges, we propose \oursys that constructs high-confidence samples named core points to obtain the model fingerprint. Through experimental analysis, we first have three insights: 1) the higher the predicted score of a sample, the farther it is from the model decision boundary, 2) piracy models output scores closer than the homologous models on core points, and 3) the farther the sample is from the model decision boundary, the greater the output score difference between homologous models and piracy models. Core points have more similar predicted scores between piracy models and the victim model due to the similarity of their decision boundaries. 
So we can identify Homologous and Piracy Models by constructing such high-confidence samples.
To solve the black-box limitation, we provide three identification methods of piracy models, including $L_1$\_dist-based, Cos\_dist-based, and clustering-based methods to measure the model outputs' correlation. 
For query efficiency, \oursys constructs at most one core point for each classification category of the victim model.
The main contributions of our work are summarized as follows:

\begin{itemize}
\item We propose a simple but novel method called \oursys to construct a model fingerprint, which can effectively and efficiently identify piracy models without misidentifying homologous models.

\item We are the first to discover the behavioral discrepancies on high-confidence classified samples between piracy and homologous models.
We have derived three insights through experimental analysis, and we utilize these insights to construct such high-confident samples. 
Finally, we design different model identification methods using these high-confidence samples.

\item Extensive experimental results demonstrate the effectiveness of \oursys in identifying various piracy models across different architectures and datasets. 
Specifically, \oursys can achieve a missed identification rate ($MIR$) and a false identification rate ($FIR$) of $0$ for piracy models, surpassing the performance of state-of-the-art methods.
\end{itemize}

\section{Related Work}
The production of piracy models poses a serious threat to the legitimate rights and interests of model owners. These models can be broadly categorized into the following types: (1) Fine-tuning \cite{tajbakhsh2016convolutional}. (2) Pruning \cite{liu2018fine,molchanov2019importance,guan2022few}. (3) Adversarial training \cite{madry2017towards}. (4) Model extraction attack \cite{jagielski2020high,orekondy2019knockoff}.
Many methods have emerged recently to protect the copyrights of model owners, which can be roughly divided into two categories: model watermarking methods and model fingerprinting methods.
(1) Model watermarking methods \cite{jia2021entangled,uchida2017embedding,fan2019rethinking,zhang2020passport,adi2018turning,ge2021anti,10059007,10636196} need to modify the parameters of the victim model and embed the watermarks carefully designed by the defender in the model, and finally verify whether it is a piracy model by detecting the watermarks. However, most watermarking methods are not effective against model extraction attacks. VEF~\cite{li2022defending} can survive during the
model extraction, but it needs white-box access to the adversary's model. In addition, the watermarking methods can also cause the loss of accuracy \cite{fan2021deepip,ge2021anti}.
(2) Model fingerprint methods \cite{lukas2019deep,cao2021ipguard,li2021modeldiff,peng2022fingerprinting,guan2022you,yangmetafinger,10201933} do not need to modify the victim model. Instead, these methods determine if a suspected model is a piracy model by analyzing its output behavior using a carefully constructed fingerprint set. 
Many existing methods \cite{lukas2019deep,cao2021ipguard,li2021modeldiff,peng2022fingerprinting} utilize adversarial examples to measure the similarity of decision boundary between the suspected model and the victim model, determining whether the suspected model is a piracy model. 
However, these methods rely heavily on adversarial examples and are not robust against adversarial defense mechanisms \cite{madry2017towards}. While MetaFinger~\cite{yangmetafinger} fingerprints the inner decision space of the model by meta-training instead of using decision boundaries. However, it does not consider homologous models and piracy models obtained by model extraction attacks.
SAC \cite{guan2022you} proposes a novel model stealing identification method based on sample correlation but does not address the false identification of homologous models. 
Inspired by the ideas of MetaFinger and SAC, we propose \oursys, a novel approach that utilizes robust samples from the inner decision space strongly associated with the victim model as the fingerprint. Our method can effectively identify between different types of piracy models and homologous models.

\section{Framework}

\subsection{Model Definition}

\textbf{Definition 1} (Victim Model).
We denote the victim model training data as $X_v\subset\{(x,y)|x\in [0,1]^M, x\sim\mu, y\in Y=\{1,\dots,N\}\}$, where $y$ represents the true label and $\mu$ represents the data distribution. We denote the victim model trained by $X_v$ as $f_v:[0,1]^M\to R^N$, aiming to optimize: $P_{(x,y)\sim X_v}(\arg\max_i f_v(x)_i = y )$. 

\textbf{Definition 2} (Homologous Model).
Homologous models denoted as $f_h:[0,1]^M\to R^N$ are legally trained by others. The training data is denoted as $X_h\subset\{(x,y)|x\in [0,1]^M, x\sim\mu, y\in Y\}$. The overlap ratio between $X_h$ and  $X_v$ can be defined by $overlap(X_h,X_v)=\frac{|X_h\cap X_v|}{|X_v|}$. The model architecture can be different from the victim model. Aiming to optimize: $P_{(x,y)\sim X_h}(\arg\max_i f_h(x)_i = y )$. 

\textbf{Definition 3} (Piracy Model).
Piracy models obtained by the adversary can be denoted as $f_p$ or $f_{\hat{p}}:[0,1]^M\to R^N$. We define $X_p\subset \{(x,y)|x\in [0,1]^M, x\sim\mu, y\in Y\}$ as the adversary's attack dataset. For piracy models obtained by illegal acquisition of the victim model, the piracy model is defined by $f_p = \Phi_{X_p}(f_v)$, where $\Phi$ represents a post-processing operation, such as fine-tuning, pruning, and adversarial training. For model extraction attacks, the adversary aims to optimize the piracy model $f_{\hat{p}}$ as follows: $P_{(x,y)\sim X_p}\left(\arg\max_i f_{\hat{p}}(x)_i = \arg\max_i f_v(x)_i \right)$.

\subsection{Threat Model}
\begin{figure}[htb]
  \centering
  \includegraphics[width=0.48\textwidth]{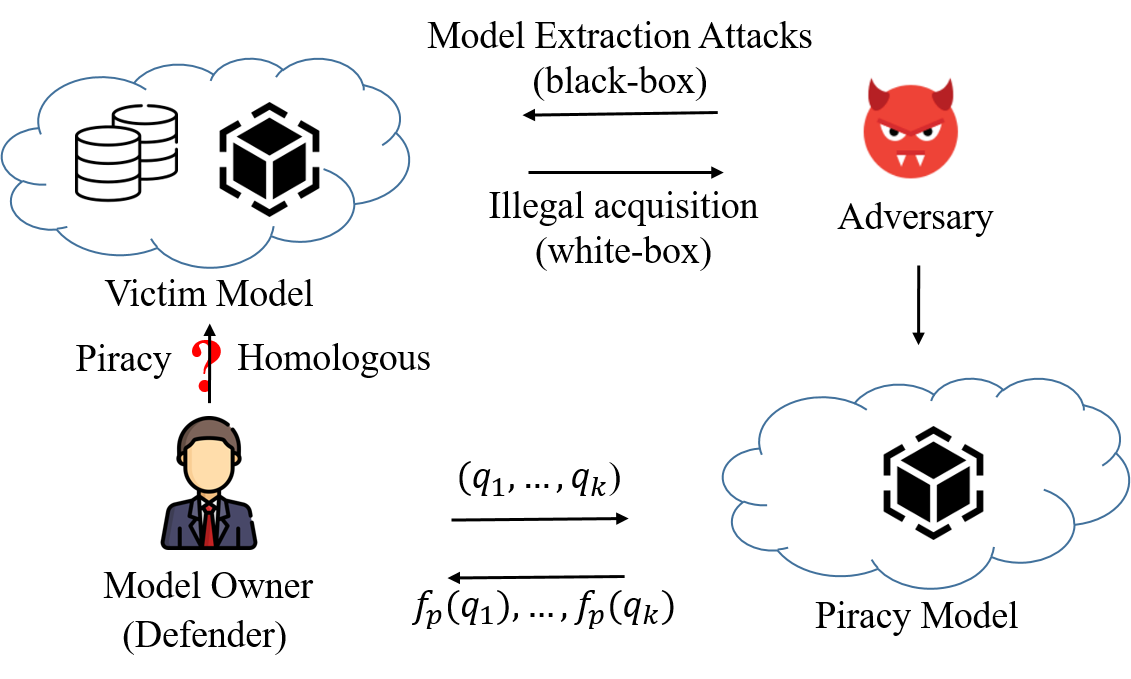}
  \caption{ A threat model of copyright protection of deep models.}
  \label{framework}
\end{figure}

Figure~\ref{framework} gives a threat model. The threat model consists of two main entities: an adversary and a defender, who is also the model owner. The adversary can obtain the defender's model in two ways: model extraction attacks and illegally acquiring the white-box model. Then the adversary could use post-processing techniques to modify the model such as fine-tuning, pruning, and adversarial training, aiming to monetize the model by just providing the model API.
Thus, the defender only has black-box access to the adversary's model, denoted as $f_p$. This means that the defender can only obtain the output of the adversary's model, denoted as $f_p(q_i)$ when providing a query sample $q_i$. The defender's goal is to identify whether it is a piracy model by analyzing the outputs of the adversary's model. 

\subsection{\oursys Design}
  \begin{figure*}[htbp]
  \centering
  \includegraphics[width=\textwidth]{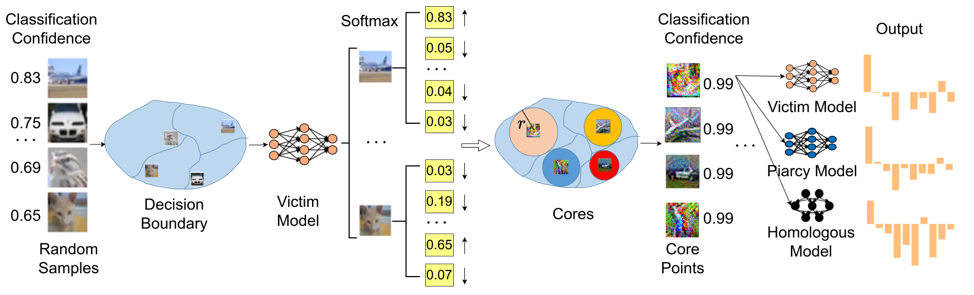}
  \caption{Given random samples as initial core points, these samples may be near the model decision boundary. For a sample, \oursys constrains its softmax score so that its classification confidence in a certain category continues to increase and in other categories continues to decrease. Finally, core points far from the decision boundary can be obtained, and the output of the piracy model for these core points is more similar to the victim model than the homologous model.}
  \label{design}
\end{figure*}

Figure \ref{design} gives the \oursys design.
The core idea of \oursys is to construct samples strongly related to the victim model's classification confidence. \oursys is based on three insights, which are analyzed in section~\ref{Insights}. For each category of the model outputs, \oursys can build high-confidence samples. Due to the behavioral discrepancies on
high-confidence classified samples between piracy and ho-
mologous models, piracy models have higher classification confidence for such samples, and the output scores are closer to the victim model than homologous models. 
Here, the score means the model's last-layer logit of the corresponding label, and the confidence means post-softmax probability.
\oursys aims to optimize the i-th core point denoted as $\phi_i$ of the corresponding label $i$ by the following loss:
 \begin{equation}
\begin{aligned}
\text{loss} = -\log \sigma(f_v(\phi_i))_i,
\end{aligned}
 \label{3.1}
\end{equation}
where $\sigma: \mathbb{R}^N\to (0,1)^N$ is a standard softmax function and is defined by the formula $\sigma(z)_i = \frac{e^{z_i}}{\sum_{j=1}^N e^{z_j}}$, for $i=1,\dots, N$, $z=(z_1,\dots,z_N)\in \mathbb{R}^N$.
Given a label $i$, the goal is to make the victim model's prediction confidence of label $i$ as high as possible.
 
Algorithm~\ref{alg1} gives the details on generating the cores.
\oursys defines $B_{r_i}(\phi_i)$ as the victim model core, where $r_i$ is the core radius. 
To initialize the core point $\phi_i$ for each label $i\in Y$, \oursys randomly selects a sample from the distribution $\mu$. Next, \oursys reduces the loss~(\ref{3.1}) by gradient descent to update the core point. 
Lines 6-16 of the algorithm are used to calculate the shortest distance from the current core point to the model decision boundary, which is defined as the core radius.
To calculate the radius of the core, DeepFool~\cite{moosavi2016deepfool} is leveraged, which can effectively compute the minimum perturbation that causes misclassification in deep neural networks.  
Through this method, for the core point $\hat{\phi}_j$ we are optimizing, we continuously iterate by adding noise $delta_j$ until the model's classification results change ($l\neq b$). Throughout this process, all noise vectors are accumulated, and their norm $r_i = ||\sum_j\delta_j||_2$ is calculated to determine the current core radius. 
Then $\Delta$ is used to calculate the difference in core radius before and after updating the core points.
The algorithm terminates when each core radius converges (i.e., $\Delta<\gamma$, where $\gamma\to 0$).
Finally, the model fingerprint denoted as $F$ is defined by $F=\{\phi_i|i=1,\dots,N\}$.
We can improve query efficiency by further reducing the number of core points with large core radii as the fingerprint.

 \begin{algorithm}
	\renewcommand{\algorithmicrequire}{\textbf{Require:}}
	\caption{\oursys}
	\label{alg1} 
	\KwIn{The victim model $f_v$, hyper-parameter $\theta$, $\gamma$} 
    \KwOut{ Cores \{$B_{r_i}(\phi_i)\}$} 
     
	\begin{algorithmic}[1]
        \FOR{Every label $i$ $\in$ Y}
		\STATE Initialize $\phi_i$, $r$, $\Delta$
		\WHILE{$\Delta>\gamma$}
            \STATE $L =  -\log \sigma(f_v(\phi_i))_i$ 
            \STATE update $\phi_i = \phi_i - \theta \nabla L $ 
            \STATE  Initialize  $j\leftarrow 0$, $\hat{\phi}_j\leftarrow\phi_i$
            \WHILE{$a:\arg\max_k f_v(\hat{\phi}_j)_k=b:\arg\max_k f_v(\phi_i)_k$}
                \FOR{$l\neq b$}
                    \STATE $\omega_l= \nabla f_v(\hat{\phi}_j)_l - \nabla f_v(\hat{\phi}_j)_b$
                \ENDFOR
            \STATE   $\hat{l} = \arg\min_{l\neq b}\frac{|f_v(\hat{\phi}_j)_l - f_v(\hat{\phi}_j)_b|}{||\omega_l||_2}$
            \STATE $\delta_j = \frac{|f_v(\hat{\phi}_j)_{\hat{l}} - f_v(\hat{\phi}_j)_b|}{||\omega_{\hat{l}}||_2}$
            \STATE $\hat{\phi}_{j+1} = \hat{\phi}_j+\delta_j$
            \STATE $j\leftarrow j+1$
            \ENDWHILE
            \STATE  $r_i = ||\sum_j\delta_j||_2$
            \STATE $\Delta = |r_i-r|$
            \STATE $r \leftarrow r_i$

        \ENDWHILE
        \ENDFOR
        \STATE return $\{B_{r_i}(\phi_i)\}\leftarrow \{\phi_i,r_i|i\in Y\}$
	\end{algorithmic}  
\end{algorithm}

\subsection{\oursys Identification}

We provide three methods to identify homologous models and piracy models. 
\oursys constructs core points by optimizing the confidence of samples in specific categories, making the core points predicted by the piracy model have scores closer to the victim model in the corresponding categories.
Therefore, the first method uses the $L_1$ distance between a victim and a suspected model to identify. Given a suspected model $f_s:[0,1]^M\to R^N$, we define the $L_1$ model distance on the model fingerprint $F$ as follows: 
$$L_1\_dist(f_v,f_s,F) = \sum_{i= 1}^N|f_v(\phi_i)_i- f_s(\phi_i)_i|.$$ 
Note that the core points constructed by \oursys can make the score of a specific category much higher than that of other categories. Each core point in the fingerprint set is very different; the second method utilizes this difference in categories to identify piracy models. 
To capture the correlations between the suspected model's outputs on the model fingerprint $F$, we use cosine similarity, denoted as $Cos(f_s, F)$ \cite{nguyen2010cosine,guan2022you} to express as follows: 
$$Cos(f_s,F)_{ij}=\frac{f_s(\phi_i)^Tf_s(\phi_j)}{||f_s(\phi_i)|| ||f_s(\phi_j)||}.$$
We define the model distance of cosine similarity as follows: 
$$Cos\_dist(f_v,f_s,F) = \frac{||Cos(f_v,F)-Cos(f_s,F)||_1}{N^2}.$$
By empirically stetting discriminant thresholds $d_1$, $d_2$, the suspected model is a piracy model if $L_1\_dist(f_v,f_s,F)<d_1$ or $Cos\_dist(f_v,f_s,F)<d_2$.
In the experiments, to determine the thresholds $d_1$ and $d_2$, we statistically analyzed the distribution of L1 distances and cosine similarities of various existing models to select suitable thresholds, aiming to maximize the differentiation between homologous and piracy models.

However, the first two threshold methods will no longer apply to identify the piracy model types. We use a clustering idea to solve this challenge. For example, we want to successfully identify homologous models, post-processing piracy models, and piracy models by model extraction attacks. 
Assume post-processing piracy models, denoted as $f_p$, piracy models obtained by model extraction attacks, denoted as $f_{\hat{p}}$, and homologous models,denoted as $f_h$,  respectively satisfy the distribution $\pi_p$,  $\pi_{\hat{p}}$ and $\pi_h$. 
we can get three cluster centers denoted as $\{\hat{c_1},\hat{c_2},\hat{c_3}\}$ from set $\{o_h,o_p,o_{\hat{p}}|f_p\sim \pi_p, f_{\hat{p}}\sim \pi_{\hat{p}}, f_h\sim \pi_h \}$, where $o_j = (f_j(\phi_1),\dots,f_j(\phi_N))$, $j\in\{h,p,\hat{p}\}$.
We use the fingerprint set as the input to query the suspected model to get an output denoted by $o_s=(f_s(\phi_1),\dots,f_s(\phi_N))$ and compare the output with the cluster centers to determine which class the suspected model belongs to.

\section{Experiments}

\subsection{Setup}

\subsubsection{Dataset} We use CIFAR-10 and CIFAR-100~\cite{2009Learning} as base datasets. CIFAR-10 comprises 60,000 images of size 32$\times$32 pixels, evenly distributed into ten categories. CIFAR-100 has 100 classes containing 600 images each. There are 500 training images and 100 testing images per class. 

\subsubsection{Victim and Homologous Model Training Data}
We divide the datasets into three parts, in which the victim model training data, the homologous model training data, and the adversary's attack data ratio is 2:2:1. We set the overlap ratio between the homologous model training set and the victim model training set from 0 to 1 and the interval is 0.1.

\subsubsection{Piracy Model Attack Methods} We conduct experimental evaluations on the following piracy models:
\begin{itemize}
\item \textbf{Fine-tuning \cite{tajbakhsh2016convolutional}}: There are two commonly used fine-tuning methods. One is to use the adversary's dataset to fine-tune only the last layer before the model output, and the other is to fine-tune all the model layers.
\item \textbf{Pruning}: The pruning strategy adopted in the experiment is Fine Pruning \cite{liu2018fine}, a combination of pruning and fine-tuning, which shows that it successfully weakens or even eliminates the model backdoor.
\item
\textbf{Adversarial Training \cite{madry2017towards}}: To evade the traditional fingerprint method of adversarial samples, the adversary introduces adversarial training to evade identification. 
\item
\textbf{Model Extraction Attacks}: We leverage the label-based model extraction attacks \cite{jagielski2020high,orekondy2019knockoff} and probability-based model extraction attacks \cite{jagielski2020high,truong2021data,gou2021knowledge}.
\end{itemize} 

\subsubsection{Homologous Model and Piracy Model Setting} All the piracy and homologous models are trained on ResNet \cite{he2016deep}, DenseNet \cite{huang2017densely}, and we use the ResNet model as the victim model architecture. 
\begin{table*}[htbp]
\newcommand{\tabincell}[2]{\begin{tabular}{@{}#1@{}}#2\end{tabular}}
\renewcommand\arraystretch{1.1}
\centering
\caption{The type and number of models.}
\resizebox{\textwidth}{!}{
\begin{tabular}{p{1cm}<{\centering}|p{2cm}<{\centering}|p{2cm}<{\centering}|p{2cm}<{\centering}|p{2cm}<{\centering}|p{2.1cm}<{\centering}}
\hline
Type  & HM\_SA  & HM\_DA & PM\_P & PM\_FL  & PM\_FA \\ 
\hline
Number & 10 & 11 & 10 & 10 & 10 \\ 
\hline
Type & PM\_Adv & EM\_SA\_L & EM\_DA\_L & EM\_SA\_Pr & EM\_DA\_Pr\\
\hline
Number & 10 & 10 & 10 & 10 & 10\\
\hline
\end{tabular}
}
\label{table_models}
\end{table*}

Table \ref{table_models} shows the type and number of test models. HM\_SA represents a homologous model trained using the same model architecture as the victim model, HM\_DA represents a homologous model trained using a different model architecture, PM\_P represents a piracy model obtained by pruning, PM\_FL represents a piracy model obtained by Fine-tuning the last layer of the victim model, PM\_FA represents a piracy model obtained by Fine-tuning all the layers of the victim model, PM\_Adv represents a piracy model obtained by adversarial training, EM\_SA\_L represents a piracy model obtained by label-based model extraction attacks with the same model architecture as the victim model, EM\_DA\_L represents a piracy model obtained by label-based model extraction attacks with a different model architecture, EM\_SA\_Pr represents a piracy model obtained by probability-based model extraction attacks with the same model architecture as the victim model, EM\_DA\_Pr represents a piracy model obtained by probability-based model extraction attacks with a different model architecture.

\subsubsection{Model IP Protection Baselines} To validate our method’s performance, we compare it with three state-of-the-art works:
\begin{itemize}
\item
SAC-w, SAC-m~\cite{guan2022you}: SAC-w selects wrongly classified samples as model inputs and calculates the mean correlation among their model outputs. SAC-m selects cut-mix augmented samples as model inputs without training the surrogate models or generating adversarial examples. 
\item
MetaFinger~\cite{yangmetafinger}:
MetaFinger proposes a robust fingerprint method about the inner decision space of the model by meta-training. 
\item
FUAP~\cite{peng2022fingerprinting}: 
FUAP proposes a novel and practical mechanism to construct fingerprints by Universal Adversarial Perturbations (UAPs).
\end{itemize} 

\subsubsection{Evaluation Metrics} To verify the effectiveness of different fingerprint methods, we have two metrics: 
\begin{itemize}
\item
Missed Identification Rate ($MIR$): The missed identification rate refers to the ratio of the number of undetected piracy models to the number of all piracy models.
\item
False Identification Rate ($FIR$): The false identification rate refers to the ratio of the number of homologous models detected as piracy models to the number of all homologous models.
\end{itemize}  

\subsection{Insight Analysis}
\begin{figure}[htbp]
  \centering
  \includegraphics[width=\linewidth]{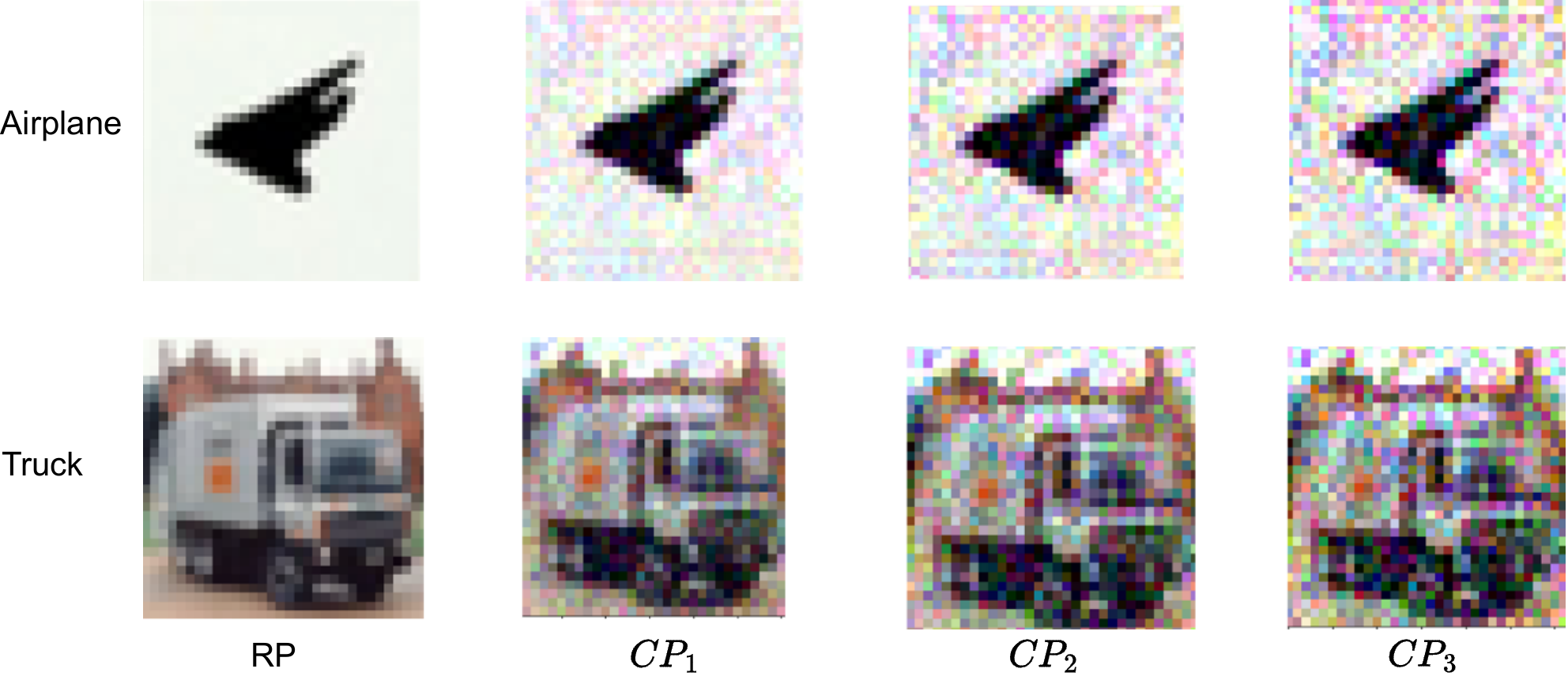}
  \caption{The visualization of the core points on CIFAR-10. The first row is an airplane. The second row is a truck. $RP$ represents the sample randomly selected as an initial core point, and $CP_i$ represents a core point trained for $i\times100$ epochs. }
  \label{rf_final}
\end{figure}

\begin{table*}[htbp]
\newcommand{\tabincell}[2]{\begin{tabular}{@{}#1@{}}#2\end{tabular}}
\renewcommand\arraystretch{1.3}
\setlength\tabcolsep{3pt}
\centering
\caption{The core radii and scores of different core points on CIFAR-10.}
\resizebox{\textwidth}{!}{
\begin{tabular}{ccccccccccc}
\hline
\bf{Radius | score}  & $r_0$| $s_0$ & $r_1$| $s_1$  & $r_2$| $s_2$ & $r_3$| $s_3$ & $r_4$| $s_4$  & $r_5$| $s_5$ & $r_6$| $s_6$ & $r_7$| $s_7$  & $r_8$| $s_8$ & $r_9$| $s_9$ \\
\hline
 Core$_1$  & 9.83|23.15  & 16.94|43.05 & 14.11|35.27 & 9.89|31.70 & 10.67|33.05 & 15.55|38.14 & 12.19|32.84 & 25.31|46.34 & 14.17|34.79 & 15.50|40.32 \\ 
 Core$_2$  & 13.61|29.06 & 23.51|56.58 & 20.24|42.35 & 11.97|37.52 & 13.64|38.74 & 18.02|46.31 & 14.82|38.05 & 28.88|54.63 & 18.77|44.79 & 18.56|46.79 \\ 
 Core$_3$  & 15.41|32.29 & 26.97|67.39 & 21.98|45.92 & 13.46|41.07 & 15.88|41.79 & 22.59|52.24 & 13.87|41.16 & 31.10|59.88 & 25.70|52.20 & 20.32|50.43 \\ 
\hline
\end{tabular}
}
\label{table_frf_R}
\end{table*}

\begin{figure*}[htbp]
  \centering
  \includegraphics[width=\linewidth]{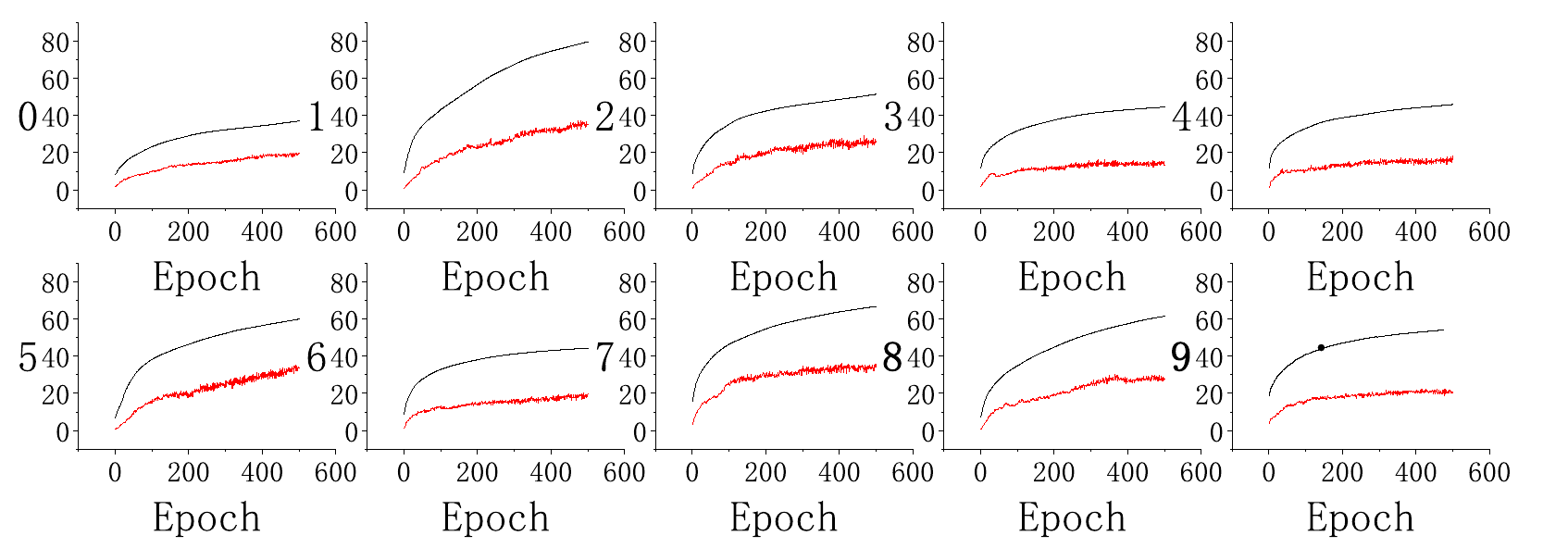}
  \caption{The black line represents the core point's score of the victim model, and the red line represents the distance from the core point to the model decision boundary. 0-9 represents the label of the current sample.}
  \label{confidence_radv}
\end{figure*}
~\label{Insights}
\subsubsection{Insight 1: The higher the core point's  predicted score, the farther the core point is from the model decision boundary}
~\label{Insight1}

Table~\ref{table_frf_R} shows the core radii and scores of different core points on CIFAR-10. Core$_i$ represents the cores with core points trained for $i\times100$ epochs. $s_i$ represents the core point's  score corresponding to label $i$, denoted as $f_v(\phi_i)_i$.
As the number of training epochs increases, the core radius of each category grows together with the score.
Figure~\ref{confidence_radv} shows that ten core points' scores of the victim model are positively correlated with the core radii within 500 training epochs.
These indicate that the higher the confidence of the core point, the further away the core point is from the model decision boundary.
Figure~\ref{rf_final} gives the visualization of the core points on CIFAR-10. The more training epochs, the more pixels are changed and the more obvious they are, which leads to poor visual effects. 

\subsubsection{Insight 2: The piracy models output scores closer to the victim model than the homologous models on core points}
~\label{Insight2}
\begin{figure*}[htbp]
  \centering
  \subfigure[A comparison of different models' scores for random samples.]{
    \label{compare_diff,subfig1} 
  \includegraphics[width=0.45\linewidth]{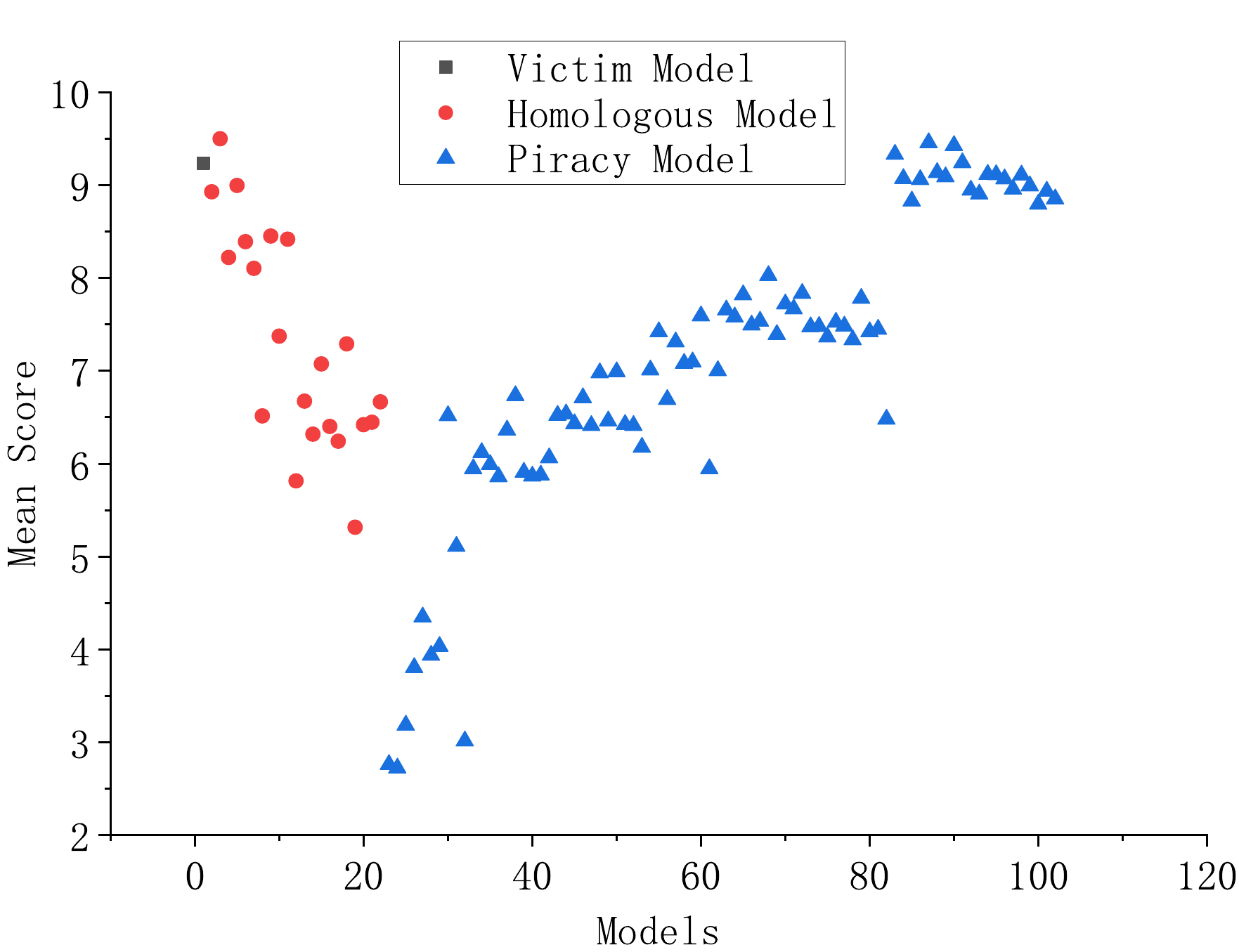}
  }
   \subfigure[A comparison of different models' scores for core points.]{
    \label{compare_diff,subfig2} 
  \includegraphics[width=0.45\linewidth]{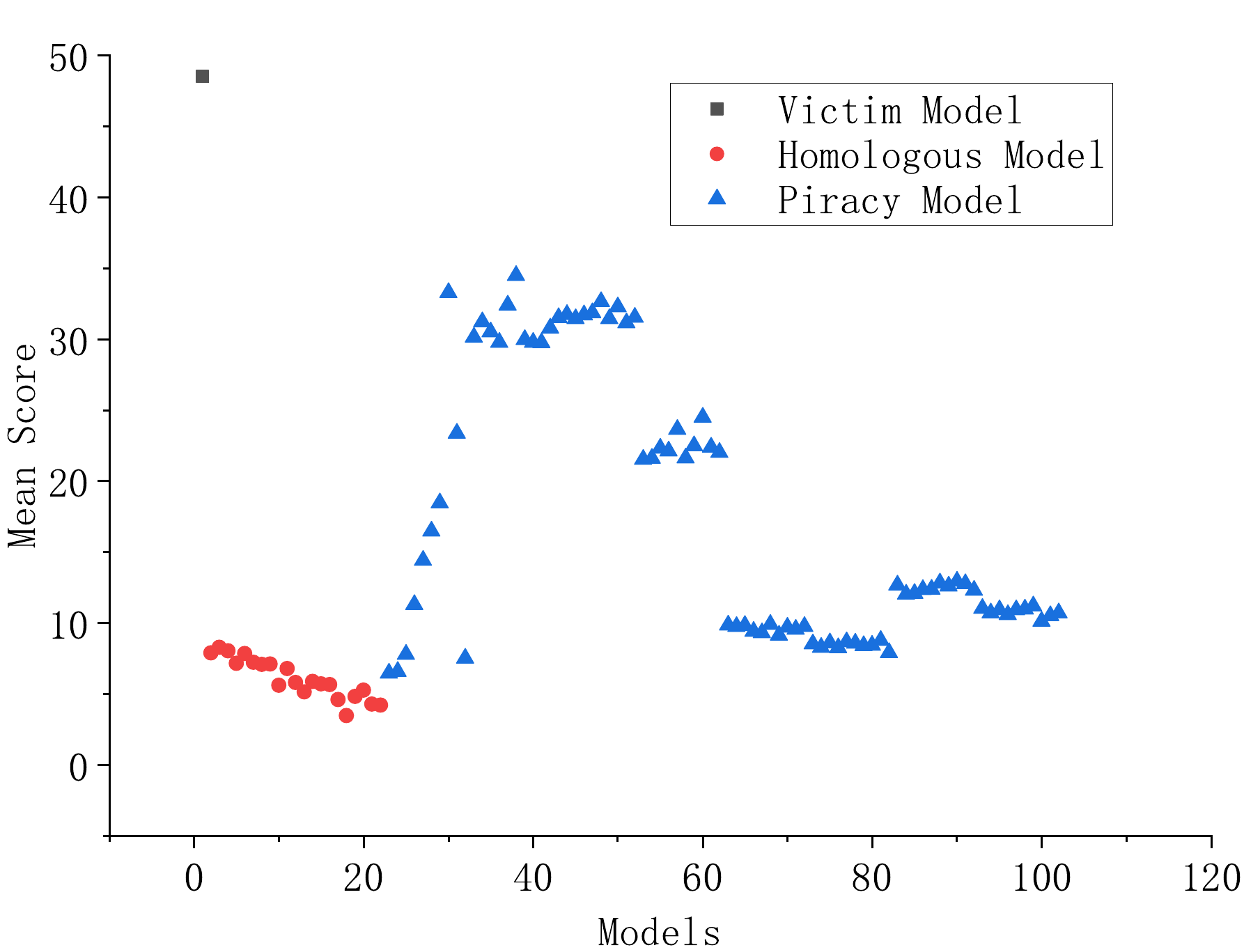}
  }
  \caption{ A comparison of different models' scores for the random samples and core points $\phi_i$ on CIFAR-10. The x-axis presents different models. The y-axis represents the average prediction score of the model on different samples.}
 
  \label{compare_diff}
\end{figure*}
Figure~\ref{compare_diff} shows different models' output scores for different samples.
There are 102 models in the figure. The first model is the victim model. Then, according to the order in table~\ref{table_models} order, the 2nd to the 22nd represent the homologous models, and numbers 23 to 102 represent various piracy models. The piracy models are more sensitive to the core points of the victim model, and most score differences between piracy models and the victim model are more minor. However, for the sample randomly selected, it is difficult to identify the homologous and piracy models from the score difference.

\begin{table*}[htbp]
\newcommand{\tabincell}[2]{\begin{tabular}{@{}#1@{}}#2\end{tabular}}
\renewcommand\arraystretch{1.1}
\centering
\caption{The average score difference between different models and the victim model.}
\resizebox{\textwidth}{!}{
\begin{tabular}{lcccccccccc}
\hline
Label  & 0  &  1 &  2 &  3 & 4& 5  & 6 &  7 &  8 & 9\\ 
\hline
RP\_h & \makecell[c]{-1.53 $\pm$2.44} & \makecell[c]{0.93$\pm$1.72} & \makecell[c]{-0.69$\pm$2.85} & \makecell[c]{5.08$\pm$1.94} & \makecell[c]{4.39$\pm$2.40} & \makecell[c]{0.13$\pm$1.41} & \makecell[c]{0.55$\pm$2.80} & \makecell[c]{5.95$\pm$2.56} &\makecell[c]{-1.68$\pm$2.26} &\makecell[c]{5.79$\pm$ 2.42}\\ 
RP\_p & \makecell[c]{0.29$\pm$1.95} & \makecell[c]{1.02$\pm$1.54} & \makecell[c]{1.30$\pm$2.07} & \makecell[c]{3.66$\pm$1.94} & \makecell[c]{3.13$\pm$1.88} & \makecell[c]{1.93$\pm$1.36} & \makecell[c]{1.00$\pm$1.98} & \makecell[c]{5.47$\pm$2.41} & \makecell[c]{-1.01$\pm$1.78} & \makecell[c]{4.72$\pm$2.59}\\
CP$_1$\_h & \makecell[c]{15.43$\pm$2.14} & \makecell[c]{32.96$\pm$2.08} & \makecell[c]{27.51$\pm$1.87} & \makecell[c]{27.44$\pm$2.34} & \makecell[c]{25.62$\pm$2.51} & \makecell[c]{32.84$\pm$2.18} & \makecell[c]{21.46$\pm$2.54} & \makecell[c]{35.55$\pm$2.71} & \makecell[c]{26.49$\pm$1.92} & \makecell[c]{28.97$\pm$2.53}\\
CP$_1$\_p & \makecell[c]{12.86$\pm$3.78} & \makecell[c]{24.36$\pm$8.76} & \makecell[c]{23.29$\pm$5.68} & \makecell[c]{20.02$\pm$3.87} & \makecell[c]{20.18$\pm$4.95} & \makecell[c]{25.76$\pm$6.77} & \makecell[c]{19.08$\pm$5.08} & \makecell[c]{28.36$\pm$7.02} & \makecell[c]{19.19$\pm$7.15} & \makecell[c]{21.29$\pm$6.16}\\
CP$_2$\_h & \makecell[c]{22.83$\pm$2.23} & \makecell[c]{47.34$\pm$2.71} & \makecell[c]{34.64$\pm$1.80} & \makecell[c]{35.20$\pm$2.73} & \makecell[c]{32.43$\pm$2.61} & \makecell[c]{42.71$\pm$2.47} & \makecell[c]{26.54$\pm$2.31} & \makecell[c]{44.89$\pm$3.39} & \makecell[c]{39.36$\pm$2.10} & \makecell[c]{36.76$\pm$2.70}\\
CP$_2$\_p   & \makecell[c]{17.72$\pm$5.33} & \makecell[c]{33.06$\pm$12.95} & \makecell[c]{28.87$\pm$7.46} & \makecell[c]{24.84$\pm$5.30} & \makecell[c]{24.59$\pm$6.47} & \makecell[c]{32.00$\pm$8.79} & \makecell[c]{22.94$\pm$6.26} & \makecell[c]{34.32$\pm$9.00} & \makecell[c]{26.77$\pm$10.69} & \makecell[c]{26.17$\pm$7.92}\\
CP$_3$\_h & \makecell[c]{26.96$\pm$2.49} & \makecell[c]{58.87$\pm$3.36} & \makecell[c]{38.67$\pm$1.92} & \makecell[c]{39.18$\pm$2.99} & \makecell[c]{36.98$\pm$2.22} & \makecell[c]{49.71$\pm$2.96} & \makecell[c]{30.79$\pm$2.29} & \makecell[c]{51.66$\pm$3.59} & \makecell[c]{49.11$\pm$2.50} & \makecell[c]{40.92$\pm$2.82}\\
CP$_3$\_p & \makecell[c]{20.29$\pm$6.30} & \makecell[c]{40.59$\pm$16.50} & \makecell[c]{31.71$\pm$8.52} & \makecell[c]{27.87$\pm$6.32} & \makecell[c]{27.40$\pm$7.64} & \makecell[c]{36.96$\pm$10.61} & \makecell[c]{25.58$\pm$7.06} & \makecell[c]{38.05$\pm$10.35} & \makecell[c]{32.74$\pm$13.23} & \makecell[c]{29.01$\pm$9.06}\\
\hline
\end{tabular}
}
\label{table_score}
\end{table*}

\begin{figure*}[htbp]
  \centering
  \subfigure[The black triangle symbol represents the label score difference between the victim model and homologous models and the red triangle symbol represents the score difference between the victim model and piracy models.]{
    \label{score,subfig1} 
  \includegraphics[width=0.6\linewidth]{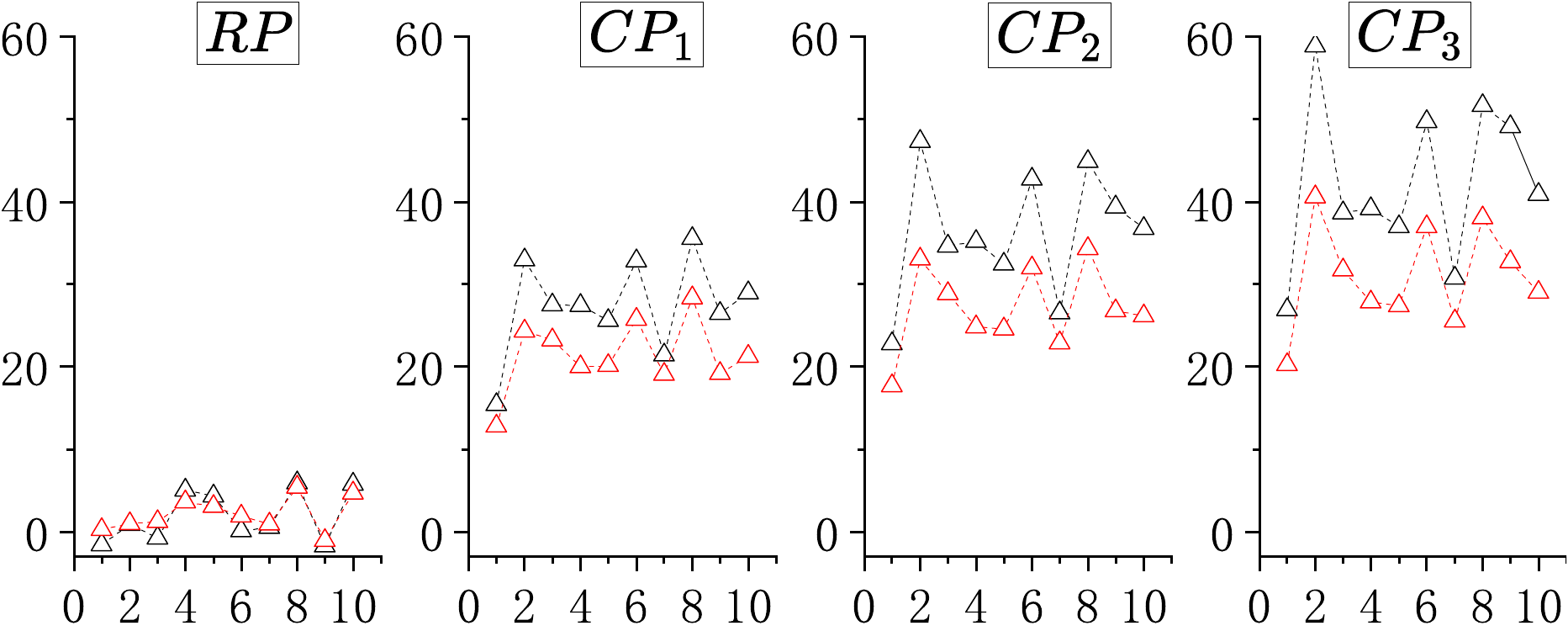}
  }
   \subfigure[The average score difference between homologous models and piracy models.]{
    \label{score,subfig2} 
  \includegraphics[width=0.33\linewidth]{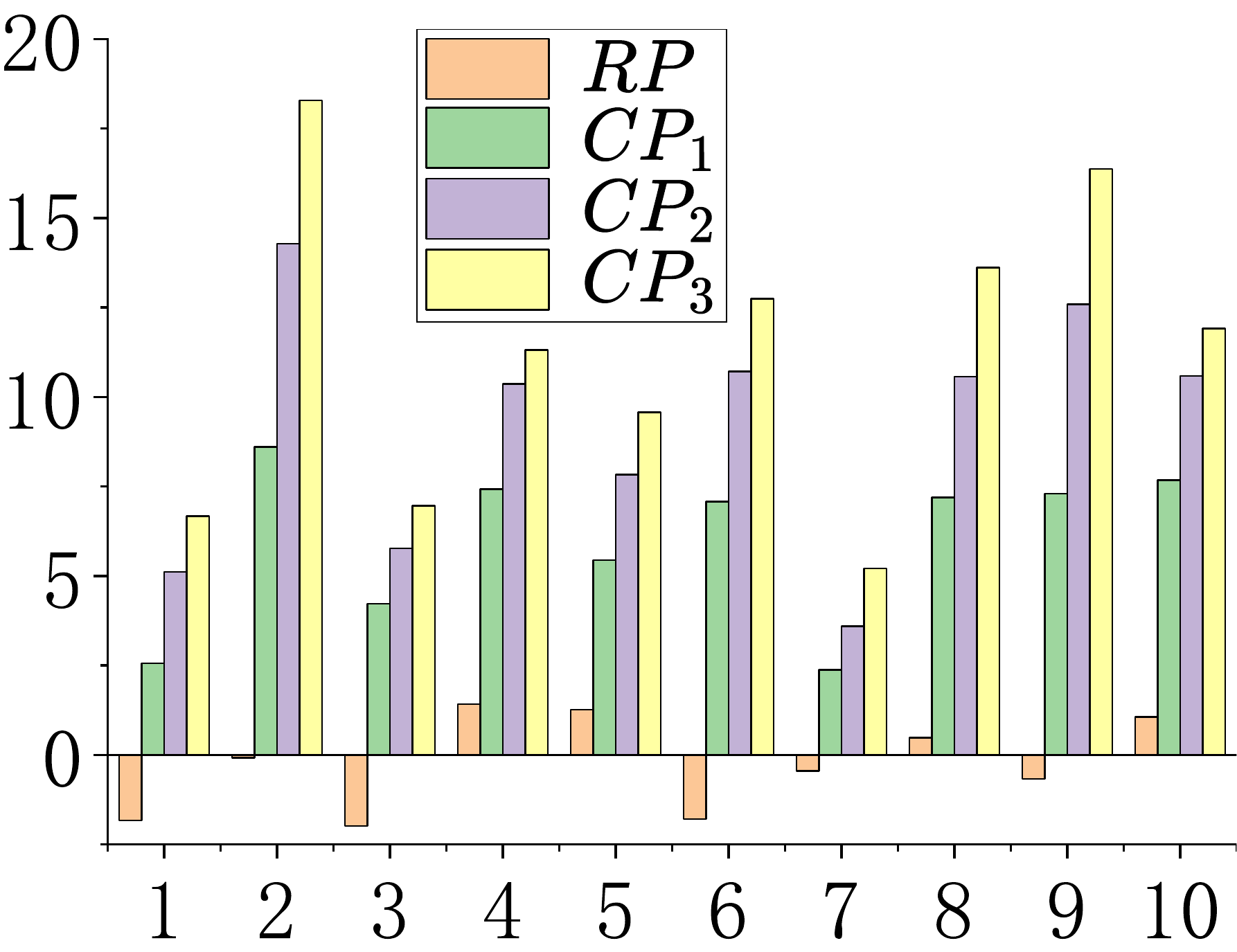}
  }
  \caption{ The x-axis in the figure represents ten core points. The y-axis represents the label score difference.}
 
  \label{score}
\end{figure*}

\begin{figure*}[htbp]
  \centering
  \subfigure[Random samples]{
    \label{ablation,subfig1} 
  \includegraphics[width=0.45\textwidth]{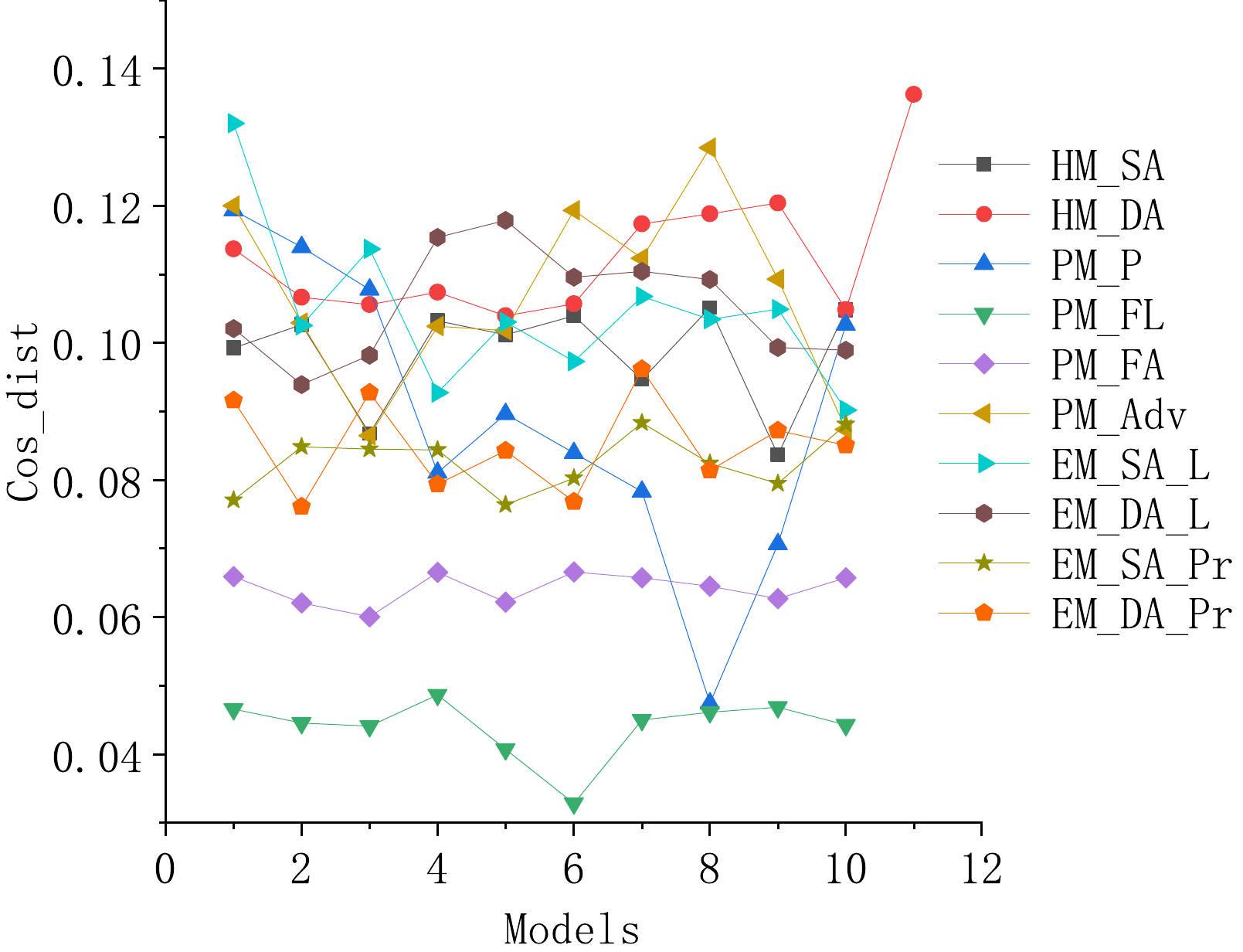}
  }
   \subfigure[Random samples]{
    \label{ablation,subfig2} 
  \includegraphics[width=0.45\textwidth]{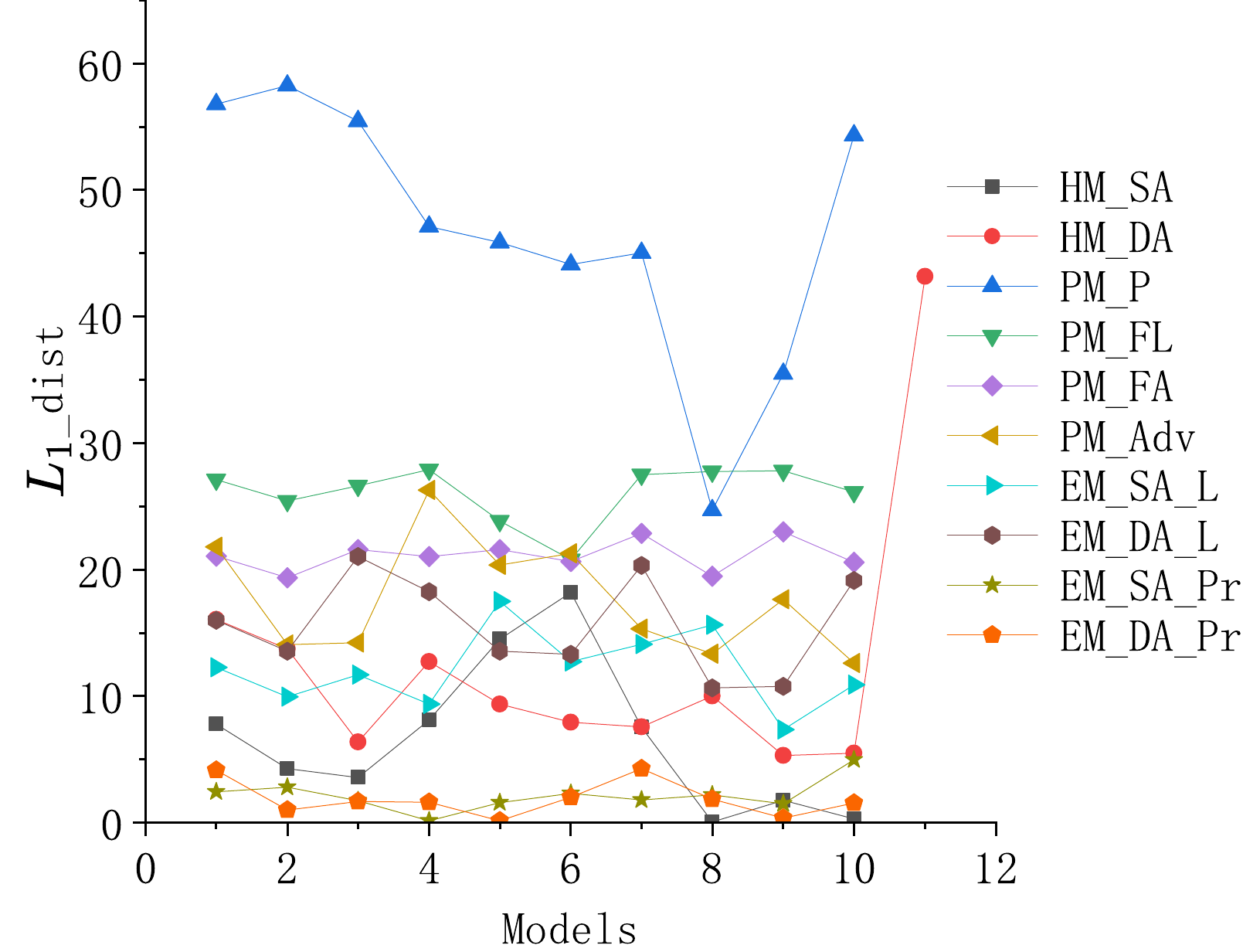}
  }
   \subfigure[\oursys]{
    \label{ablation,subfig3} 
  \includegraphics[width=0.45\textwidth]{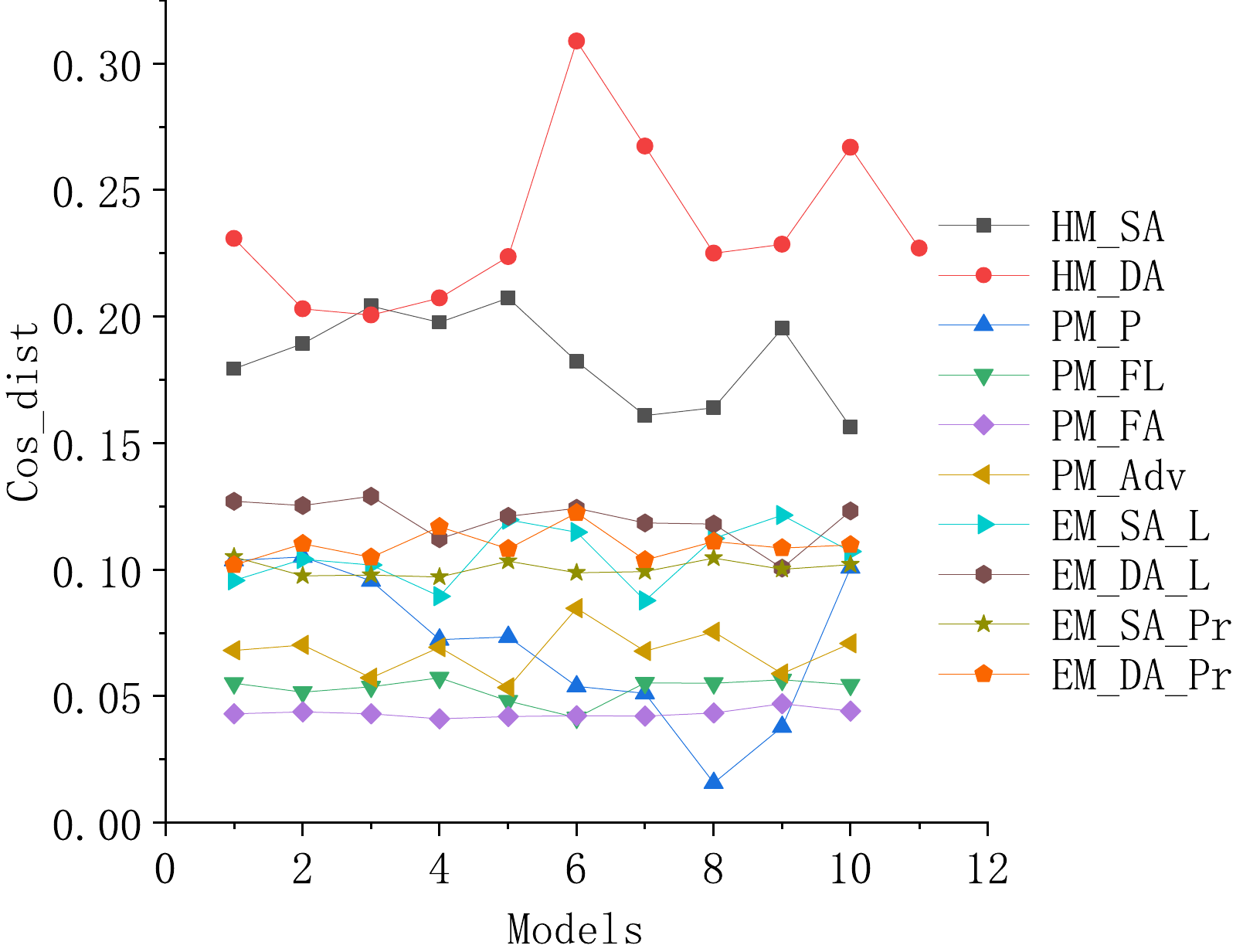}
  }
   \subfigure[\oursys]{
    \label{ablation,subfig4} 
  \includegraphics[width=0.45\textwidth]{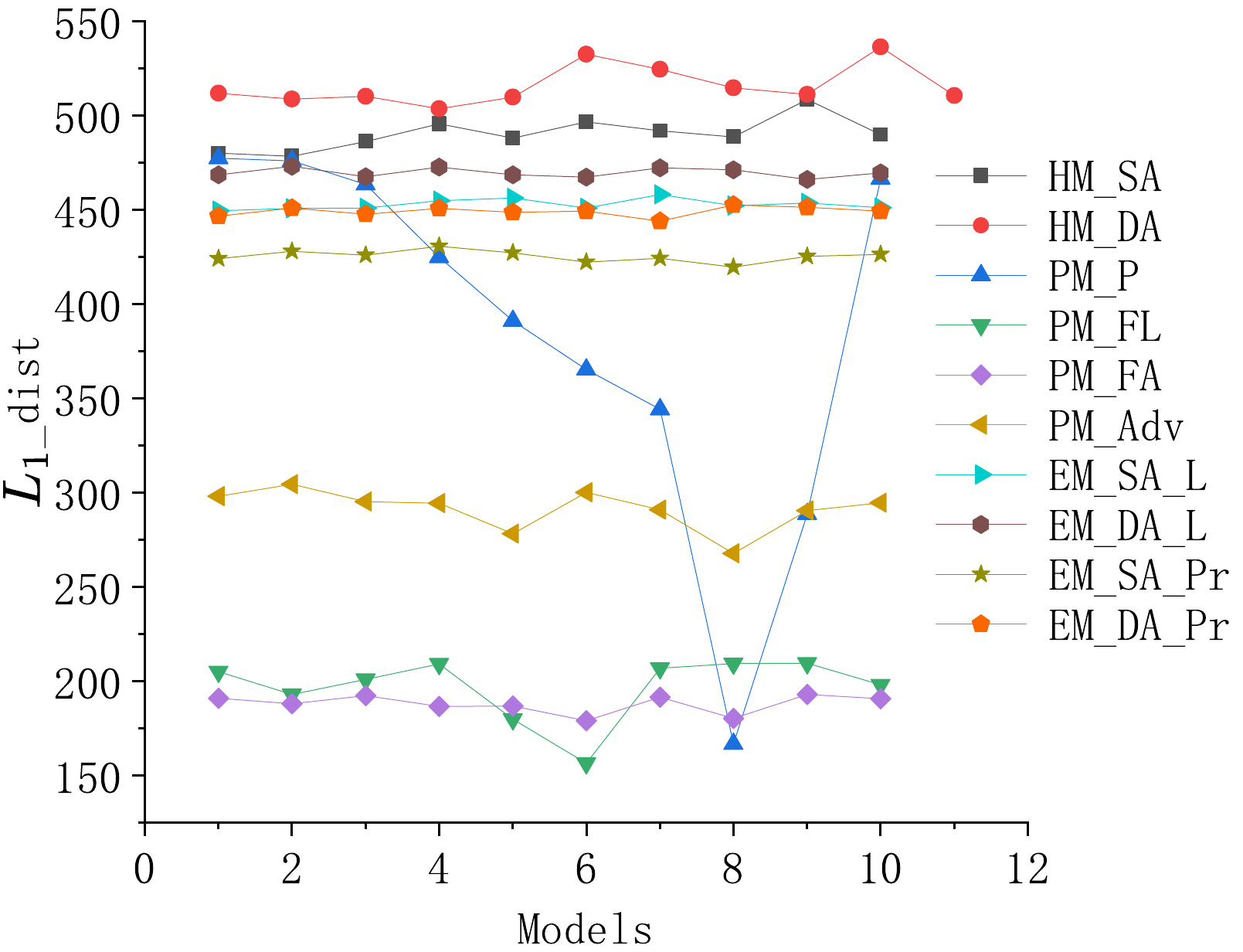}
  }
  \caption{The effectiveness of the random samples' optimization of \oursys.}  
  \label{ablation}
\end{figure*}

\subsubsection{Insight 3: The larger the core point's radius, the greater the core point's score difference between homologous models and piracy models}
In table~\ref{table_score}, we calculate the average score difference and the variance between different models and the victim model. A label score difference between a suspected model and the victim model is denoted as $f_v(\phi_i)_i-f_s(\phi_i)_i$ for the core point $\phi_i$. RP\_h represents an initial core point's score difference between a homologous model and the victim model. RP\_p represents such a difference between a piracy model and the victim model. 
CP$_i$ represents a core point trained for i × 100 epochs. 
CP$_i$\_h represents  CP$_i$'s score difference between a homologous model and the victim model. CP$_i$\_p represents  CP$_i$'s score difference between a piracy model and the victim model. 0 to 9, respectively, represent the 10 RP or CP$_i$ of the corresponding label $i$.
Combined with figure~\ref{score}, we can draw the following conclusions:
The larger the core radius, the greater the score difference between homologous models and piracy models. 
Figure~\ref{confidence_radv} shows that the larger the epoch, the larger the core radius.
Therefore, training core points for more epochs makes it easier to identify piracy models.

\subsection{\oursys Performance}

\subsubsection{The effectiveness of \oursys}

Figure \ref{ablation} shows the effectiveness of \oursys compared to initial random samples. If we use random samples as fingerprints, it is difficult to use the threshold to separate the homologous models from the piracy models.

\subsubsection{Comparative experiments}

Table \ref{table_fingerprinting_methods} gives the performance comparison between \oursys and baselines. It can be seen that the \oursys has the best $MIR$ and  $FIR$.
Moreover, \oursys can perform well under three different identification methods.

\begin{table}[htbp]
\newcommand{\tabincell}[2]{\begin{tabular}{@{}#1@{}}#2\end{tabular}}
\renewcommand\arraystretch{1.1}
\centering
\caption{Comparative  experiments on CIFAR-10 and CIFAR-100.}
\resizebox{0.5\textwidth}{!}{
\begin{tabular}{p{3.3cm}<{\centering}p{1cm}<{\centering}p{1cm}<{\centering}p{1cm}<{\centering}p{1cm}<{\centering}}
\hline
\multirow{2}*{\bf{Method }} &\multicolumn{2}{c}{CIFAR-10} &\multicolumn{2}{c}{CIFAR-100}\\
 & \bf{$MIR\downarrow$} & \bf{$FIR\downarrow$} & \bf{$MIR\downarrow$} & \bf{$FIR\downarrow$} \\
\hline
SAC-w~\cite{guan2022you}   & 0.05 & 0.20 & 0.23 & 0.43\\ 
SAC-m~\cite{guan2022you}  & 0.21 & 0.10 & 0.61 & 0.52\\ 
FUAP~\cite{peng2022fingerprinting}  & 0.04 & 0.05 & 0.06 & 0.05\\ 
MetaFinger~\cite{yangmetafinger}  & 0.07 & 0.65 & 0.14 & 0.94\\ 
\oursys $(L_1\_dist)$  & 0.03 & 0.00 & 0.09 & 0.05\\ 
\oursys $(Cos\_dist)$   & \bf{0.00} & \bf{0.00} & \bf{0.03} & \bf{0.04}\\ 
\oursys $Clustering$  & \bf{0.00} & \bf{0.00} & 0.00 & 0.09\\ 
\hline

\end{tabular}
}
\label{table_fingerprinting_methods}
\end{table}

\subsubsection{The impact of different clustering methods}
\begin{figure*}[htbp]
  \centering
  \subfigure[KMeans]{
    \label{cluster,subfig1} 
  \includegraphics[width=0.3\textwidth]{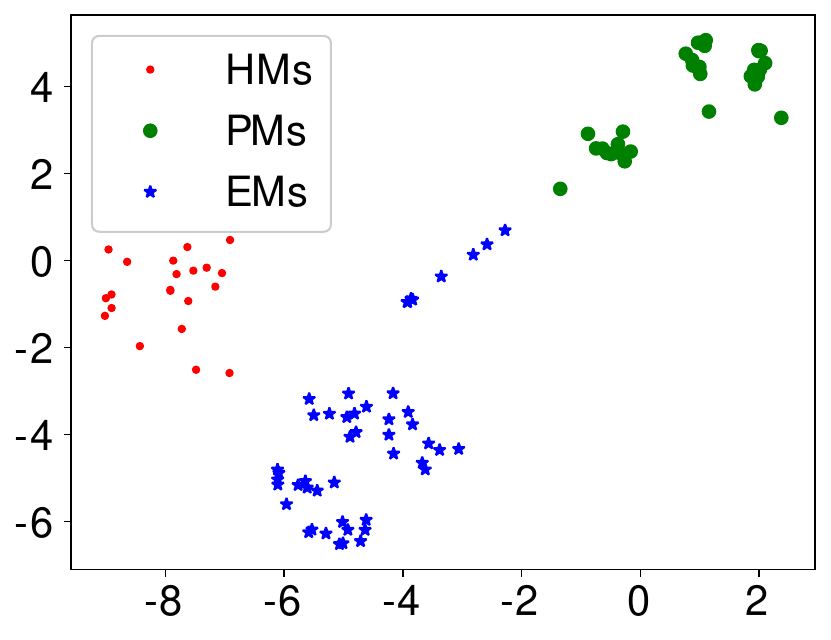}
  }
   \subfigure[SpectralClustering]{
    \label{cluster,subfig2} 
  \includegraphics[width=0.3\textwidth]{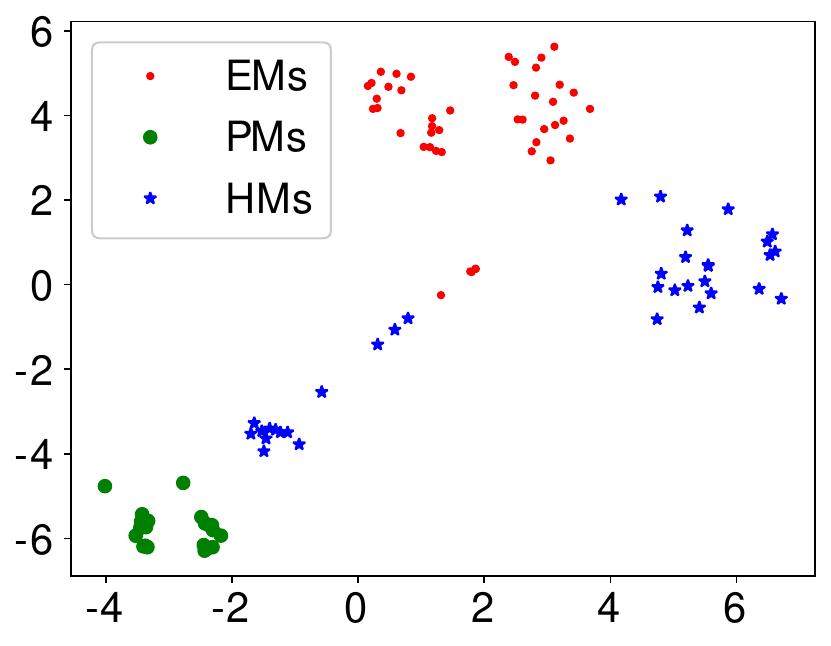}
  }
  \subfigure[AgglomerativeClustering]{
    \label{cluster,subfig3} 
  \includegraphics[width=0.3\textwidth]{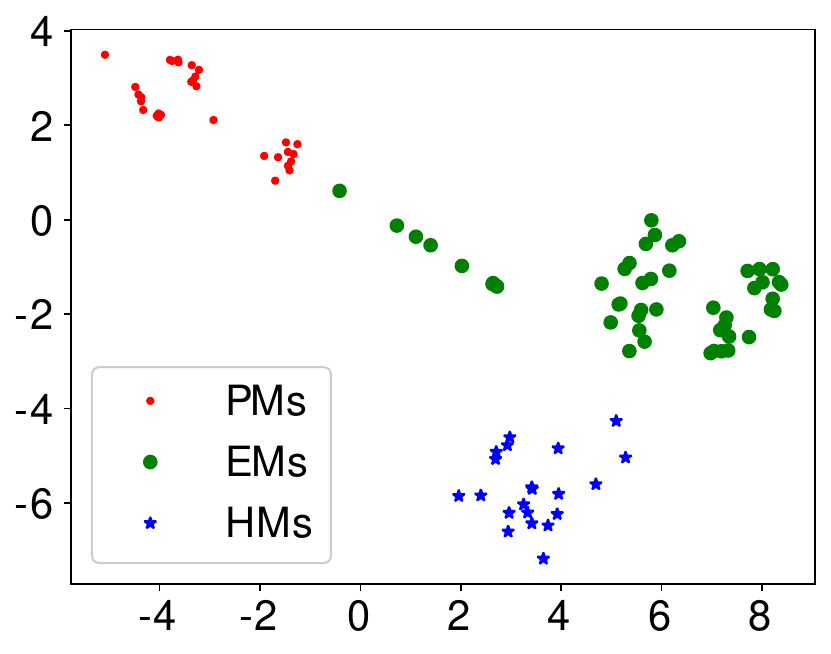}
  }
  \caption{The visualization of different clustering methods.}  
  \label{cluster}
\end{figure*}

In the \oursys identification stage, we can use clustering methods to identify piracy models at a more fine-grained level. 
In our experiments, we use Kmeans~\cite{macqueen1967some}, SpectralClustering (SC)~\cite{ng2001spectral}, and AgglomerativeClustering (AC)~\cite{dasgupta2005performance}.  
\begin{table}[htbp]
\newcommand{\tabincell}[2]{\begin{tabular}{@{}#1@{}}#2\end{tabular}}
\renewcommand\arraystretch{1.1}
\centering
\caption{The experiment results of clustering methods.}
\begin{tabular}{ cccc } 
\hline
\bf{Clustering Methods}  & \bf{KMeans}  & \bf{SC} & \bf{AC}\\
\hline
$MIR$ (PMs) & 0.20 & 0.48 & 0.23\\ 
$FIR$ (PMs)& 0.00 & 0.00 & 0.00\\ 
$MIR$ (EMs) & 0.00 & 0.00 & 0.00\\ 
$FIR$ (EMs)& 0.17 & 0.11 & 0.20\\ 
$MIR$ (HMs) & 0.00 & 0.00 & 0.00\\ 
$FIR$ (HMs)& 0.00 & 0.40 & 0.00\\ 
\hline
\end{tabular}
\label{table_clustering_methods}
\end{table}

Table \ref{table_clustering_methods} shows the $MIR$ and $FIR$ of PMs, EMs, and HMs for different clustering methods. 
PMs refer to models that have undergone post-processing techniques. EMs represent piracy models acquired through model extraction attacks. HMs represent homologous models. Kmeans and AC perform better than SC, which shows that different clustering methods also affect identification accuracy. 
Figure \ref{cluster} intuitively shows the SpectralClustering results of HMs are very scattered and can mistakenly identify piracy models. Although Kmeans and AgglomerativeClustering cannot wholly distinguish different piracy models, they can at least distinguish homologous models well. 
Therefore, choosing a suitable classification algorithm is also crucial and worthwhile to continue exploring.
The clustering-based method can identify piracy models in a more fine-grained manner, but the disadvantage is that the number of types of piracy models needs to be known in advance.

\section{Discussion}
In our approach, we need to empirically set a threshold or the number of clusters to identify piracy models. Therefore, similar to the previous methods, it is still necessary to set the hyperparameters based on the prior statistical results. 
Another limitation is that
the samples constructed by our method have poor visual effects, so \oursys has a potential flaw, that is, it can be discovered by the suspected model owner during the query. 
To solve the poor visual effects, similar to adversary examples, we can set the upper limit of pixel variation. It is a balance between visual effects and the difficulty of fingerprint search. 
Thirdly, the minimum number of core points required for a model fingerprint is also an interesting question.
Finally, in the era when large models are getting bigger and bigger, how to extend our work to large models is a very meaningful issue.
We will leave these problems to future work. 

\section{Conclusion}

This work proposes a simple fingerprint construction framework called \oursys to identify different piracy and homologous models. 
Different from previous adversarial example-based and correlation-based fingerprint methods, 
this paper takes a different approach, exploring the impact of sample points within the model decision space on the output behavioral discrepancies of different models, and derives three insights through extensive experimental analysis. These insights are then used to construct model fingerprints, enabling the differentiation of piracy and homologous models. Given the complexity and uninterpretability of deep models, we believe our findings can also contribute to the exploration of model decision spaces.

\section*{Acknowledgments}
The research is partially supported by National Key R\&D Program of China  No.2018YFB0803400, National Key R\&D Program of China under Grant No.2021ZD0110400, China National Natural Science Foundation with No.62132018, Key Research Program of Frontier Sciences, CAS. No.QYZDY-SSW-JSC002, The University Synergy Innovation Program of Anhui Province with No.GXXT-2019-024.

\bibliographystyle{IEEEtran}
\bibliography{main}

\end{document}